\documentclass [12 pt]{article}

\usepackage{a4wide}
\usepackage{exscale}
\usepackage{amsmath}
\usepackage{amsfonts}
\usepackage{amssymb}
\usepackage{epsf}
\usepackage{graphicx}
\usepackage{subfigure}
\usepackage{hyperref}

\begin{document}

\begin{center}{\bf{Novel black hole bound states and entropy}}

\bigskip

{T.R. Govindarajan\footnote{e-mail address: {\tt trg@imsc.res.in}} and Rakesh Tibrewala\footnote{e-mail address: {\tt rtibs@imsc.res.in}}}

\bigskip

{\it Insitute of Mathematical Sciences,}\\
{\it Chennai, 600 113, India}

\end{center}


\begin{abstract}

We solve for the spectrum of the Laplacian as a Hamiltonian on $\mathbb{R}^{2}-\mathbb{D}$ 
and in $\mathbb{R}^{3}-\mathbb{B}$. A self-adjointness analysis
with $\partial\mathbb{D}$ and $\partial\mathbb{B}$ as the boundary for the two cases 
shows that a general class of boundary conditions for which the Hamiltonian operator 
is essentially self-adjoint are of the mixed (Robin) type. 
With this class of boundary conditions we obtain ``bound state" solutions  
for the Schroedinger equation. Interestingly, these solutions are all localized near 
the boundary. We further show that the number of bound states is 
finite and is in fact proportional to the perimeter or area
of the removed \emph{disc} or \emph{ball}. We then argue 
that similar considerations should hold for static 
black hole backgrounds with the horizon  treated as the boundary.

\end{abstract}

\section{Introduction}

Ever since Bekenstein's suggestion that black holes are thermodynamical objects that 
have entropy
\cite{bekenstein1} and Hawking's discovery that black holes radiate 
\cite{hawking}, there have 
been numerous attempts to explain the thermodynamical properties of black holes from more 
fundamental quantum principles. The celebrated area law for black hole entropy 
suggests that the 
microscopic degrees of freedom reside on the black hole horizon and that
the number of these states is proportional to the exponential of the horizon area. It is 
generally believed that a full understanding of black hole thermodynamics would 
require knowledge of
quantum theory of gravity. Black hole entropy has been calculated in several ways - (i) brickwall (ii) string theory, and (iii) loop quantum gravity (LQG).  
It is well known in many of these formulations that states contributing to the entropy 
are obtained from a conformal field theory near the boundary 
\cite{strominger,ashtekar,kaul,carlip,paddy}. 

Although entropy calculations in string theory and LQG give results in 
agreement with what is expected from semiclassical arguments, they 
require details of quantum gravity. Nevertheless, it is also believed that, 
at least for large black holes, it must be possible to explain black hole 
entropy without requiring the details of quantum gravity. One such proposal 
is that the black hole entropy is the result of quantum entanglement 
between the degrees of freedom of a scalar field that are outside and those which are inside the 
event horizon (since the degrees of freedom inside the horizon are 
inaccessible to a distant observer), see \cite{shanki}. 

't Hooft has proposed that black hole states form a discrete spectrum 
and has shown that the number 
of energy levels that a particle can occupy in the vicinity of a black hole 
is finite if one imposes a cut-off in the form of a \emph{brick wall} near the 
horizon \cite{thooft1, thooft2}. 
Based on an earlier work \cite{bekenstein2}, Bekenstein and Mukhanov \cite{mukhanov} independently suggested that the black hole discrete mass levels 
give rise to the black hole entropy as well as the thermal radiation from it. 
These calculations, though interesting, are not illuminating 
in explaining the origin of the  discrete spectrum of energy levels. 

Taking seriously the view that entropy and Hawking radiation of large black 
holes should not 
depend critically on the details of quantum gravity, we suggest an 
alternate proposal. The main ingredient behind the new proposal 
is the existence of bound quantum states in black hole backgrounds resulting 
from the self-adjoint extension of the Laplacian on these backgrounds. 
If these bound states are localized near the horizon and if their number is finite, 
then one has a candidate model to explain black hole entropy. 
Possibility of bound state solutions in the vicinity of black hole horizon 
has been explored previously also \cite{trg}.  

Here we carry that program further but now restricting ourselves to the Schroedinger 
equation and its bound state solutions. The crucial requirement is that the boundary conditions
are determined through self-adjointness. This serves two purposes: (i) the evolution for
arbitrary combinations of modes will be unitary, and (ii) arbitrary $L^2$ functions in the domain 
of the Laplacian can be expanded in a complete set of eigenmodes.  
Interestingly, we find that most of the 
expectations mentioned in  previous paragraph are realized. 
We find that there is a finite number of discrete bound state solutions and that 
these solutions are localized near the horizon. We also find that the number 
of these bound state solutions is proportional to the horizon area. 

To present the key ideas and the intuition involved in this proposal we start, 
in Sec. \ref{toy models}, by considering toy models in 
$2+1$ and $3+1$ dimensional flat spacetimes 
with a ball removed from the spatial sections. We solve 
for the spectrum of the Laplacian and look for bound state solutions. To figure out 
the suitable domain of the operator in which it is self-adjoint  
we perform self-adjointness analysis. 
We find that the boundary conditions 
for which this operator is self-adjoint are the so-called Robin (mixed) 
boundary conditions. For the most general boundary conditions in which 
the operator is self-adjoint see \cite{ibort}. 

We find that the Hamiltonian admits bound state solutions 
in the exterior of this model black hole and that the number of these bound 
state solutions is proportional to the "area" of the horizon 
(circumference of the disc in $2+1$ dimensions). In other words, 
for these model black holes, we find that the entropy is proportional 
to the area. 

After this warm-up, in Sec. \ref{btz spacetime}, we apply similar ideas to 
the case of $2+1$ dimensional nonrotating Banados, Teitelboim and Zanelli BTZ black hole. However, 
due to the nontrivial metric and the existence of horizon (a null surface), 
the analysis now 
becomes much more subtle (see \cite{pulak} for instance). After identifying the Laplacian we proceed 
with its self-adjointness analysis (using measure appropriate for this background). 
We show explicitly that the Laplacian is self-adjoint with 
mixed Robin type boundary conditions in tortoise coordinates.  

One important feature to note here is that the use of mixed boundary conditions, 
as opposed to the usual Dirichlet or Neumann boundary conditions, naturally 
introduces a length parameter in the problem. In the present context, 
where one is taking into account both the gravitational and the quantum effects, 
it is natural to take this parameter to be of the order of the 
Planck length $l_{p}$. The Dirichlet and the Neumann boundary conditions 
turn out to be two special cases of the mixed boundary condition where 
the length parameter goes to infinity and zero, respectively. The effect of 
these boundary conditions has been studied in an analysis of 
the billiards problem in \cite{berry}.
In fact these modes in a compact region of space give phenomena like whispering gallery 
modes and are also responsible for evanescent waves.
That boundary conditions  and self-adjoint extensions are in one to one relation was
established by Asorey et. al. \cite{ibort}. We will comment on this later in the discussion.
Interesting new constraints arise when we have Dirichlet conditions at isolated points, see Berry et. al. \cite{berry}.  
It is worth noting that the occurrence of 
localized bound states in the presence of mixed boundary conditions  
have implications for edge states in the context of 
quantum hall effect as well.

\section{Toy models} \label{toy models}
We are interested in finding the bound state solutions of the 
time-independent Schroedinger equation 
\begin{equation} \label{general eigenvalue equation}
-\triangle\psi(r,\Phi)=\Lambda\psi(r,\Phi) \,.
\end{equation}
Here $\triangle$ is the Laplacian operator, $r$ is the radial 
coordinate, and $\Phi$ designates all the angular coordinates. Note that 
for bound state solutions $\Lambda<0$. The solution to the above equation 
is to be obtained subject to the Robin boundary condition
\begin{equation} \label{robin boundary condition}
\kappa\psi+\partial_{\overrightarrow{n}}\psi=0 \,,
\end{equation}
where $\kappa$ is a constant and the derivative in the second term is 
evaluated along the outward normal $\overrightarrow{n}$ to the boundary. 
It is well known that the Laplacian operator is self-adjoint in the 
domain of $L^2$ functions subject to the above condition 
\eqref{robin boundary condition}.
We note that the constant $\kappa$ has dimensions of 
$\rm{(length)^{-1}}$. This parameter would not be present if we use the 
usual Dirichlet or Neumann boundary conditions (which are the limits of 
Robin boundary condition corresponding, respectively, to $\kappa^{-1}$ or 
$\kappa\rightarrow 0$). 

\subsection{2+1 dimensional flat spacetime}
We begin by considering the case of two dimensional Euclidean plane with a disc of radius $r_{b}$ removed from it. For two spatial dimensions \eqref{general eigenvalue equation}, in terms of polar coordinates, becomes
\begin{equation} \label{eigenvalue equation in 2d}
\frac{1}{r}\frac{\partial}{\partial r}\left(r\frac{\partial\psi}{\partial r}\right)+\frac{1}{r^{2}}\frac{\partial^{2}\psi}{\partial\theta^{2}}=\lambda\psi \,,
\end{equation}
where we have made the replacement $\Lambda\rightarrow-\lambda$ so that for 
bound states $\lambda>0$. Since our boundary conditions do not mix the coordinates we can 
solve using the ansatz $\psi(r,\theta)=e^{in\theta}R(r)$ (with $n$ an integer) 
resulting in the following $r$ dependent equation:
\begin{equation} \label{r dependent equation in 2d}
\frac{d^{2}R}{dr^{2}}+\frac{1}{r}\frac{dR}{dr}-\frac{n^{2}R}{r^{2}}=\lambda R \,.
\end{equation} 
Before solving this equation we check whether the Laplacian,
\[
\hat{O}=\frac{d^{2}}{dr^{2}}+\frac{1}{r}\frac{d}{dr}-\frac{n^{2}}{r^{2}} \,,
\]
is self-adjoint \cite{reed, trg, kumargupta} on the domain 
$D(\hat{O})=\{\psi|\psi \, \rm{continuous \, and} \, 
\psi\in\mathcal{L}^{2}[r_{b},\infty), \kappa\psi(r_{b})-\psi'(r_{b})=0\}$, 
defined by the boundary condition in \eqref{robin boundary condition}. 
Here prime $(')$ denotes derivative with respect to $r$.  

We start by confirming that the operator $\hat{O}$ is symmetric, 
that is, $(\phi, \hat{O}\psi)-(\hat{O}\phi, \psi)=0 \, \, 
\forall \, \phi, \, \psi \in D(\hat{O})$ where the inner product, 
using the measure $rdr$, is
\begin{equation} \label{inner product}
(\phi, \hat{O}\psi)=\int_{r_{b}}^{\infty} r\phi^{*}\hat{O}\psi dr \,.
\end{equation}
Substituting for $\hat{O}$ we get
\begin{equation}
(\phi, \hat{O}\psi)-(\hat{O}\phi, \psi)= r_{\infty}
\left(\phi^{*}_{\infty}\frac{d\psi_{\infty}}{dr}-
\psi_{\infty}\frac{d\phi^{*}_{\infty}}{dr}\right) \,,
\end{equation}
after an integration by parts and using the boundary 
condition in the resulting expression.
If $\psi_{\infty}$ and its derivative falloff sufficiently 
fast at infinity, the above boundary term  is zero 
thereby confirming that the operator is symmetric. 

To check if the operator is self-adjoint we need 
to find the domain of the adjoint operator - 
the set of all those $\phi$'s, not necessarily belonging to 
$D(\hat{O})$, which satisfy $(\phi, \hat{O}\psi)=(\hat{O}\phi, \psi) \, 
\forall \, \psi \in D(\hat{O})$. Using the same procedure as above, 
but with $\phi$ arbitrary, we get
\begin{equation} \label{self adjointness in 2d}
r_{\infty}\phi^{*}_{\infty}\frac{d\psi_{\infty}}{dr}-\kappa r_{b}\phi^{*}_{r_{b}}\psi_{r_{b}}-r_{\infty}\psi_{\infty}\frac{d\phi^{*}_{\infty}}{dr}+r_{b}\psi_{r_{b}}\frac{d\phi^{*}_{r_{b}}}{dr}=0 \,.
\end{equation}
Since $\psi$ and its derivative fall off sufficiently fast at 
infinity, the above equation gives following condition on $\phi$ at $r_{b}$:
\begin{equation}
\kappa\phi^{*}_{r_{b}}=\frac{d\phi^{*}_{r_{b}}}{dr} \,.
\end{equation}
This implies that $\phi \in D(\hat{O})$, that is, the domain of the 
adjoint is the same as the domain of the operator itself. 
This establishes that the operator is self-adjoint with mixed boundary conditions.

\begin{figure}[htbp]
\begin{center}
\includegraphics[scale=.4]{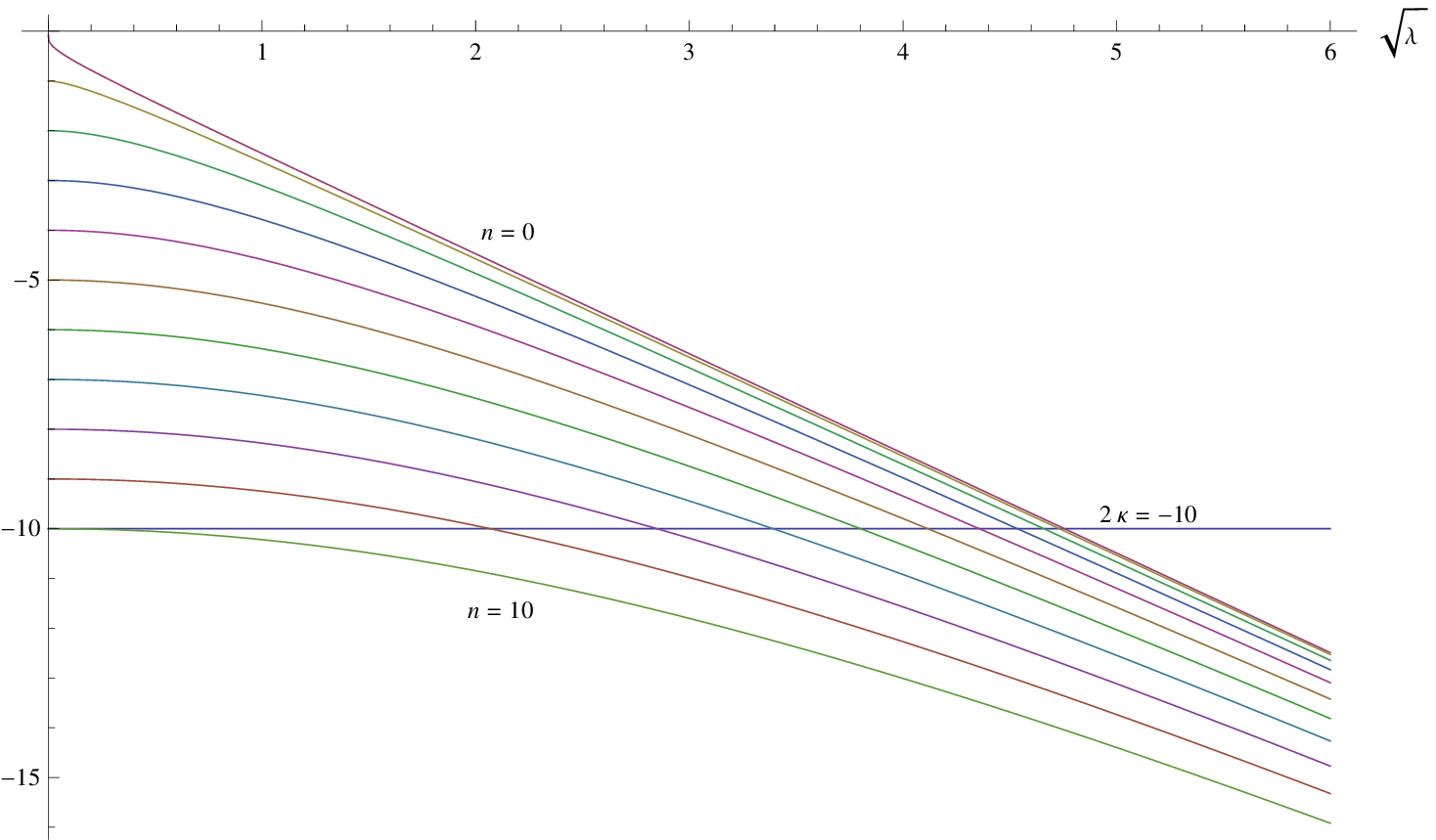}
\end{center}
\caption{\label{2d bound solutions}} Bound solutions in two dimensions for $r_{b}=2$. The horizontal line corresponds to $\kappa=-5$.
\end{figure}

Having proved the self-adjointness of the operator in \eqref{r dependent 
equation in 2d}, we now come to the solutions of that equation. These are 
given by modified Bessel's function $I_{n}(\sqrt{\lambda}r)$ and 
$K_{n}(\sqrt{\lambda}r)$. Requiring the solutions to be square integrable 
on $(r_{b},\infty)$ with measure $rdr$ ($r_{b}$ being the boundary), 
rules out the exponentially growing $I_{n}(\sqrt{\lambda}r)$ solution, and 
we are left with
\begin{equation} \label{modified bessel function}
R(r)= cK_{n}(\sqrt{\lambda}r) \,,
\end{equation}
where $c$ is a normalization constant. The complete solution is thus given by 
\begin{equation}
\psi(r,\theta)=cK_{n}(\sqrt{\lambda}r)e^{in\theta} \,.
\end{equation}
To find the spectrum, we now impose the boundary condition 
in \eqref{robin boundary condition} at $r=r_{b}$ obtaining
\begin{equation}
\kappa=\frac{\sqrt{\lambda}K'_{n}(\sqrt{\lambda}r_{b})}
{K_{n}(\sqrt{\lambda}r_{b})} \,.
\label{boundary condition in 2d}
\end{equation}
Here the sign on the right-hand side (rhs) is positive since the derivative is 
evaluated along the \emph{outward} normal which, in the present context, 
points towards the origin. 
Since $K'_{n}(r)<0$, we find that there will be bound solutions only for $\kappa<0$. 
By plotting the left-hand side (lhs) and the rhs of the above equation (for different values of $n$) 
as a function of $\sqrt{\lambda}$ we find that there are only a 
finite number of bound solutions for a given value of $\kappa$ and $r_{b}$, 
these being given by the intersection of the above mentioned curves with the 
horizontal line representing the constant value of $\kappa$, 
see Fig.~\ref{2d bound solutions}.

To find how the number of bound state solutions scale with the boundary 
radius $r_{b}$, we start by noting that the derivative of the rhs of  
\eqref{boundary condition in 2d} with respect to $\sqrt{\lambda}$ in the 
limit $\sqrt{\lambda}\rightarrow0$ and under the assumption that $n$ is 
sufficiently greater than $\sqrt{\lambda}$, tends to zero (remark: it turns out that we only need $n>2$ for the argument to work). This implies that the tangent to the curves corresponding to the rhs 
of \eqref{boundary condition in 2d} are parallel to the $\sqrt{\lambda}$ axis 
and therefore also to the $\kappa=$ constant curve. 

Thus, in the limit 
$\sqrt{\lambda}\rightarrow0$ the curve corresponding to some order 
$n_{m}$ ($>2$, see the remark above) of the modified Bessel function will either be coincident with 
the $\kappa= \rm{constant}$ curve or will be the first curve not intersecting it at all 
(in either case the solution corresponding to order $n=n_{m}$ will not 
be a bound state). Thus, one can find the dependence of the total number of 
bound state solutions on the radius $r_{b}$ of the removed disc. For this 
we take the limit $\sqrt{\lambda}\rightarrow0$ on the rhs of 
\eqref{boundary condition in 2d} with $n=n_{m}$ and use the approximation
\begin{equation} \label{approximation for modified bessel function}
K_{\alpha}(x)\approx\frac{\Gamma(\alpha)}{2}\left(\frac{2}{x}\right)^{\alpha} \quad (\sqrt{\alpha+1}\gg x>0)
\end{equation}
to obtain
\begin{equation}
\lim_{{\sqrt{\lambda}\rightarrow0}} {\sqrt{\lambda}}\frac{K_{n_{m}}'(\sqrt{\lambda}r_b)}
{K_{n_{m}}(\sqrt{\lambda}r_b)} = -\frac{n_{m}}{r_{b}} \,.
\end{equation}
Equating this to $\kappa$ one finds that $n_{m}$ scales linearly with radius $r_{b}$
\begin{equation} \label{scaling of nm with rb in 2d}
n_{m}=-\kappa r_{b} \,.
\end{equation}

From the arguments presented above, the maximum number of bound state 
solutions is equal to $n_{m}$ if the rhs of \eqref{scaling of nm with rb in 2d} 
is an integer (the total number of bound state solutions is $n_{m}$ and not 
$n_{m}-1$ since $n=0$ also corresponds to a bound state solution). If it is 
not an integer, then the number of bound state solutions is the smallest integer 
greater than $-\kappa r_{b}$. We also note that except for the case $n=0$ 
there is a two-fold degeneracy in the number of bound state solutions 
since $n\rightarrow -n$ gives the same solution. 

\begin{figure}[htbp]
\begin{center}
\begin{tabular}{cc}
\subfigure[$E_{n}~v/s~n$]{\includegraphics[scale=.25]{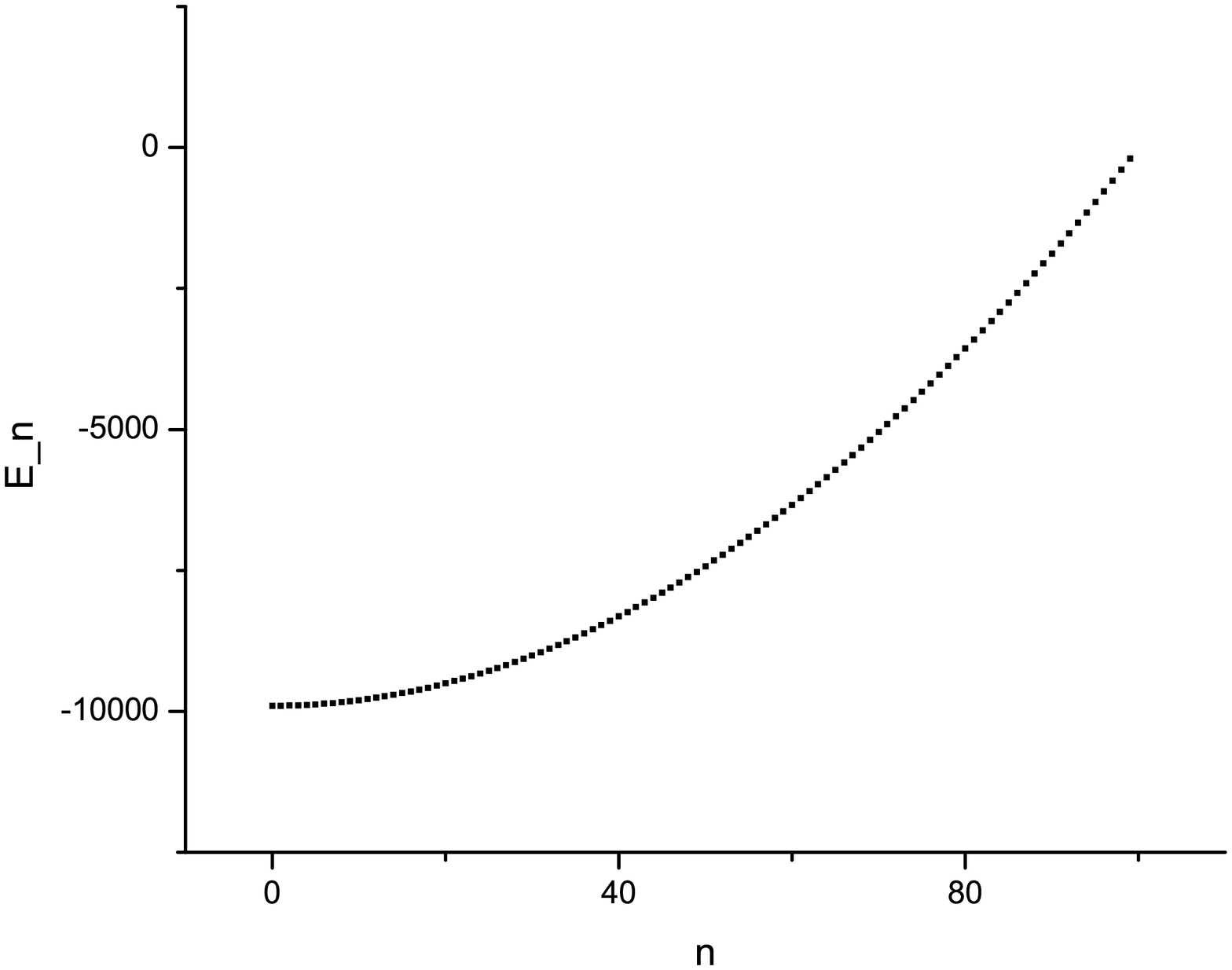}} &
\subfigure[$<r_{n}>~v/s~n$]{\includegraphics[scale=.25]{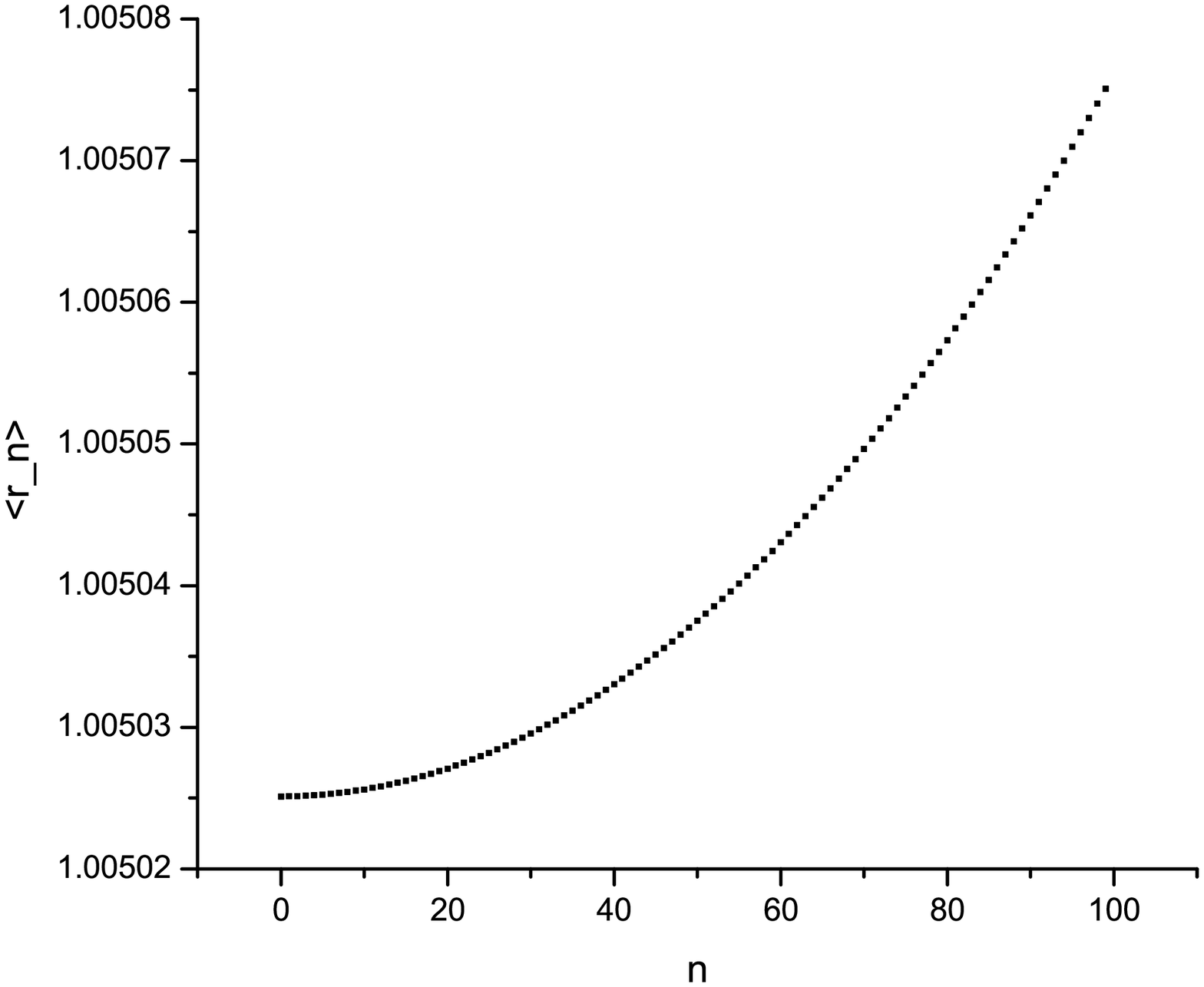}} \\
\end{tabular}
\end{center}
\caption{\label{2d solutions kappa 100 rb 1}} Energy spectrum $E_{n}$ and the expectation value $<r_{n}>$ in two spatial dimensions for $r_{b}=1$ and $\kappa=-100$.
\end{figure}



From \eqref{boundary condition in 2d} it is difficult  
to obtain an analytic expression for the 
energy of bound states. However the spectrum can be obtained numerically and in Fig.~\ref{energy spectrum 2d solutions} and \ref{expectation value of r 2d solutions}
we show the dependence of energy $E_{n}$ and the expectation value $<r_{n}>$ on $n$, the order of bound states, for the choice $\kappa=-100$ and $r_{b}=1$. 

From the figure we note that the energy of the lowest state $E_{0}$ is 
close to $-\kappa^{2}$. The spectrum is approximately given by $E_{n}\approx-\kappa^{2}+n^{2}/r_{b}^{2}$ and thus with increasing $n$ the energy approaches zero.
We also note that for all $n$ the expectation value $<r_{n}>$ is 
always close to $r_{b}$ showing that these states are all localized 
near the boundary. This is a general feature of the solution as can be verified 
by choosing different values for $\kappa$ and $r_{b}$. 

\subsection{3+1 dimensional flat spacetime}

We now repeat the above calculations for the case of $\mathbb{R}^{3}-\mathbb{B}^{3}$. Using the expression for the Laplacian in spherical coordinates, the eigenvalue problem to be solved is
\begin{equation} \label{eigenvalue equation in 3d}
\frac{1}{r^{2}}\frac{\partial}{\partial r}\left(r^{2}\frac{\partial \psi}{\partial r}\right)+\frac{1}{r^{2}\sin{\theta}}\frac{\partial}{\partial\theta}\left(\sin{\theta}\frac{\partial \psi}{\partial \theta}\right)+ \frac{1}{r^{2}\sin^{2}\theta}\frac{\partial^{2}\psi}{\partial \phi^{2}}=\lambda\psi \,,
\end{equation}
where as before $\lambda>0$. The above equation can be solved easily by using separation of variables writing $\psi(r,\theta,\phi)=R(r)Y_{lm}(\theta,\phi)$, where $Y_{lm}(\theta,\phi)$'s are the spherical harmonics solving the angular part of the equation. The radial part of the equation is given by (in subsequent analysis we replace $l$ by $n$)
\begin{equation} \label{r dependent equation in 3d}
\frac{d^{2}R}{dr^{2}}+\frac{2}{r}\frac{dR}{dr}-\frac{n(n+1)R}{r^{2}}=\lambda R \,.
\end{equation}

Since the form of the Laplacian is the same as that in \eqref{r dependent equation in 2d}, 
it is clear that this again is a self-adjoint operator on the domain of interest. 
As in the previous case the above equation, with the requirement of square integrability on $(r_{b},\infty)$ 
with measure $r^{2}dr$, is solved by the modified Bessel function of the 
second kind $K_{n+\frac{1}{2}}(\sqrt{\lambda}r)$ with the only difference that now  
the order of the modified Bessel function is $(n+1/2)$. The complete solution to the equation is
\begin{equation}
\psi(r,\theta,\phi)=c\frac{K_{n+\frac{1}{2}}(\sqrt{\lambda}r)}{r^{1/2}}Y_{nm}(\theta,\phi) \,.
\end{equation}

We now impose the boundary condition in \eqref{robin boundary condition} and as in the two dimensional case we have condition only on the radial solution
\begin{equation}
\kappa R(r_{b})-R'(r_{b})=0 \,.
\end{equation}

\begin{figure}[htbp]
\begin{center}
\includegraphics[scale=.5]{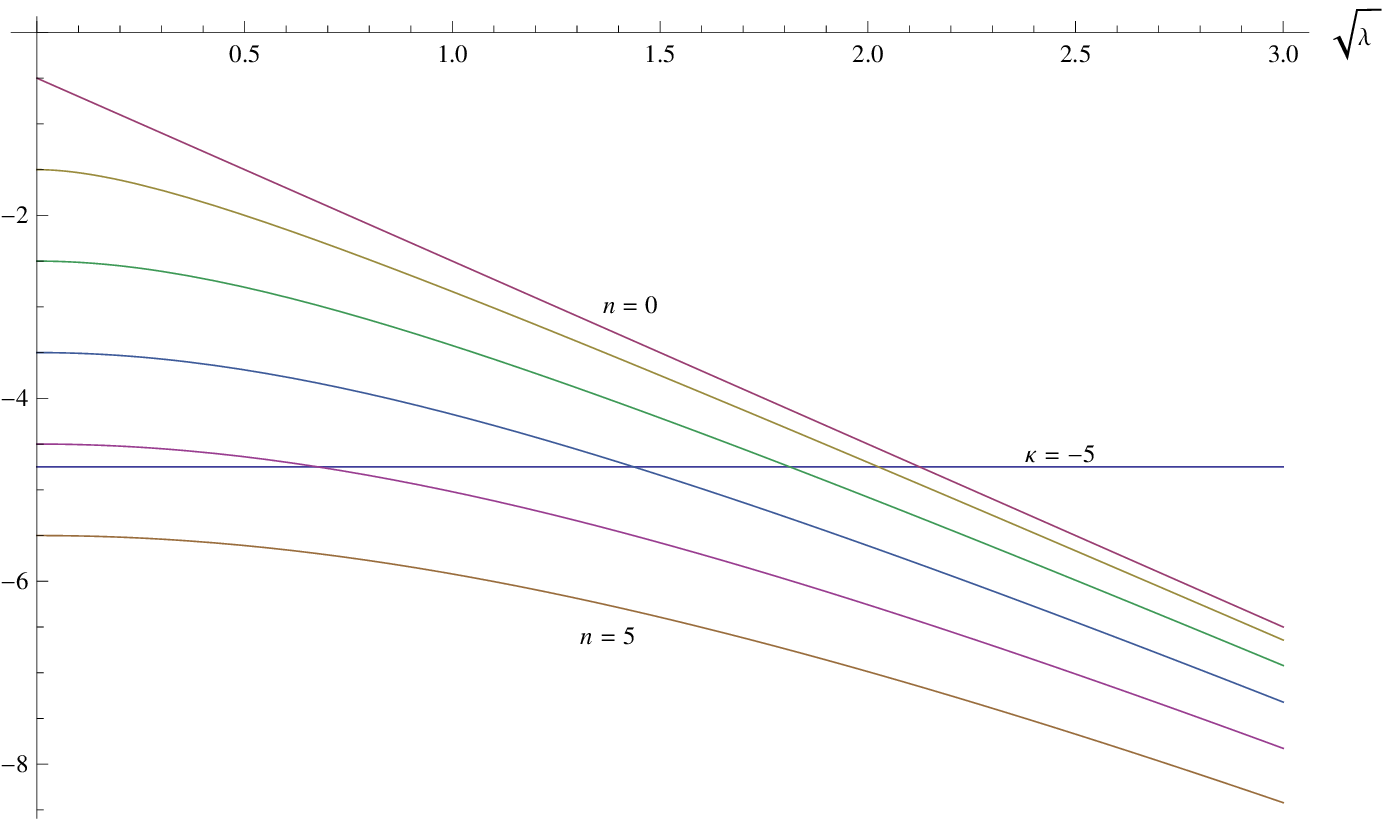}
\end{center}
\caption{\label{3d bound solutions}} Bound solutions in three dimensions for $r_{b}=2$. The horizontal line corresponds to $\kappa=-5$.
\end{figure}

After substituting for $R$ this becomes
\begin{equation} \label{boundary condition in 3d}
2\kappa=-2\sqrt{\lambda}\frac{K_{n-\frac{1}{2}}(\sqrt{\lambda}r_{b})+K_{n+\frac{3}{2}}(\sqrt{\lambda}r_{b})}{K_{n+\frac{1}{2}}(\sqrt{\lambda}r_{b})}-\frac{1}{r_{b}} \,.
\end{equation}
We again find that bound solutions are possible only for $\kappa<0$ and that for a given value of $\kappa$ and $r_{b}$ only a finite number of bound solutions are possible (see Fig.~\ref{3d bound solutions}).

We find the scaling of the total number of bound state solutions with $r_{b}$ by proceeding exactly as in the two dimensional case. Taking the limit $\sqrt{\lambda}\rightarrow0$ on the rhs of \eqref{boundary condition in 3d} and using the approximation \eqref{approximation for modified bessel function} one finds that
\begin{equation}
\lim_{\sqrt{\lambda}\rightarrow0}-2\sqrt{\lambda}\frac{K_{n_{m}-\frac{1}{2}}(\sqrt{\lambda}r_{b})+K_{n_{m}+\frac{3}{2}}(\sqrt{\lambda}r_{b})}{K_{n_{m}+\frac{1}{2}}(\sqrt{\lambda}r_{b})}-\frac{1}{r_{b}}=-\frac{(4n_{m}+3)}{r_{b}} \,.
\end{equation}
Equating this to $2\kappa$ we find that as in the two dimensional case, $n_{m}$ scales linearly with $r_{b}$ in three dimensions also
\begin{equation} \label{scaling of nm with rb in 3d}
n_{m}=-\frac{(2\kappa r_{b}+3)}{4} \,.
\end{equation}
The number of bound state solutions (ignoring the degeneracy factor) is then given by
\begin{eqnarray} \label{total number of bound state solutions in 3d}
n_{b} &=& -\frac{(2\kappa r_{b}+3)}{4}-1 \quad \mbox{($n_{m}$ an intger)} \,, \nonumber \\
n_{b} &=& \lfloor-\frac{(2\kappa r_{b}+3)}{4}\rfloor \quad \quad \mbox{($n_{m}$ not an integer)} \,.
\end{eqnarray}
Taking into account the degeneracy in the spherical harmonics, the total number of bound states is given by $\Sigma_{i=0}^{i=b-1}(2n_{i}+1)=n_{b}^{2}$. Using the expression for $n_{b}$ in \eqref{total number of bound state solutions in 3d} we see that this has terms depending quadratically and linearly on $r_{b}$ as well as a term independent of it. However for $r_{b}\gg0$, the quadratically dependent term dominates over the other two. 
 

\section{The Nonrotating BTZ Black Hole} \label{btz spacetime}

The analysis of the previous section can be done for black hole spacetimes as well, 
treating the black hole horizon as a boundary. One of the main interests would be to see 
whether similar results hold in that case as well. If they do then it would provide a 
plausible explanation for the origin of black hole entropy without requiring a detailed 
knowledge of quantum gravity.  
The picture would be a realization (at least for a large black hole)
that the bound states are localized near the horizon and their 
number (to leading order) is proportional to the area of the black hole. 

With these remarks, we now apply the methods of the previous section to the case of 
BTZ black hole \cite{btz}. To simplify calculations we limit ourselves to the case of 
nonrotating BTZ black holes. We want to obtain the Laplacian for a spatial slice in this 
background. The main point of the calculation is to do a self-adjointness analysis 
of the Laplacian and see if it implies mixed boundary condition which, as noted earlier, 
would introduce a length parameter in the problem that can be taken to be of the 
order of the Planck length. Thus we will need to identify the Laplacian operator 
on the BTZ spacetime, the metric for which (in the nonrotating case) is given by
\begin{equation} \label{btz metric}
ds^{2}=g_{\alpha\beta}dx^{\alpha}dx^{\beta}=-\frac{(r^{2}-r_{+}^{2})}{l^{2}}dt^{2}+\frac{l^{2}}{(r^{2}-r_{+}^{2})}dr^{2}+r^{2}d\theta^{2} \,,
\end{equation}
 where $r_{+}$ corresponds to the horizon and $l^{2}$ is the negative inverse of 
the cosmological constant. The massive Klein-Gordon equation is given by
\begin{equation} \label{kg eq}
(\square-m^{2})\phi=\frac{1}{\sqrt{-g}}\partial_{\alpha}(\sqrt{-g}g^{\alpha\beta}\partial_{\beta})\phi-m^{2}\phi=0 \,,
\end{equation}
where $g^{\alpha\beta}$ is the inverse of the metric in \eqref{btz metric} 
and $\sqrt{-g}$ is its determinant. Using \eqref{btz metric} in \eqref{kg eq} we get
\begin{equation} \label{kg eq in r coord}
-\frac{l^{2}}{(r^{2}-r_{+}^{2})}\partial_{t}^{2}\phi+\frac{1}{r}\partial_{r}\left[\frac{r(r^{2}-r_{+}^{2})}{l^{2}}\partial_{r}\right]\phi+\frac{1}{r^{2}}\partial_{\theta}^{2}\phi-m^{2}\phi=0 \,.
\end{equation}
This equation has been analyzed in detail at several places, see \cite{ichinose} 
for instance, and it is known that the equation can be solved exactly. 

\subsection{The measure in the $r$ coordinate}
We are interested in the self-adjointness of the
Laplacian which requires a suitable measure (defining the inner product) on the 
Hilbert space. In a general curved background the scalar product is defined by
\cite{BirrellDavies}
\begin{equation} \label{scalar product curved background}
(\phi_{1},\phi_{2})=-i\int_{\Sigma}[\phi_{1}(\partial_{\mu}\phi_{2}^{*})-(\partial_{\mu}\phi_{1})\phi_{2}^{*}]\sqrt{g_{_{\Sigma}}}n^{\mu}d\Sigma \,.
\end{equation}
Here $n^{\mu}$ is the future directed unit vector orthogonal to spacelike 
hypersurface $\Sigma$; $g_{_{\Sigma}}$ is the determinant of the metric on the
hypersurface with $d\Sigma=drd\theta$ being the volume element on it. For the 
BTZ metric, \eqref{btz metric}, the unit normal to the hypersurface is given
by $g^{\alpha\beta}n_{\alpha}n_{\beta}=g^{tt}n_{t}n_{t}=-1$. This implies that the 
only nonzero component of $n^{\mu}$ is $n^{t}=-l/(r^{2}-r_{+}^{2})^{1/2}$. We
also have $\sqrt{g_{_{\Sigma}}}=lr/(r^{2}-r_{+}^{2})^{1/2}$ and $d\Sigma=drd\theta$. 
Thus the inner product for nonrotating BTZ is given by
\begin{equation} \label{scalar product btz}
(\phi_{1},\phi_{2})=i\int_{\Sigma}[\phi_{1}(\partial_{t}\phi_{2}^{*})-(\partial_{t}\phi_{1})\phi_{2}^{*}]\frac{l^{2}r}{(r^{2}-r_{+}^{2})}drd\theta \,.
\end{equation}
Note that the measure diverges near the horizon (this is related to the redshift).

\subsection{Equations in tortoise coordinates}
It turns out that the problem can be translated in the form of standard 
Schroedinger equation by going to the tortoise coordinates. This also makes 
the analysis of self-adjointness issue easier. We use the ansatz
$\phi=e^{-i\omega t}e^{in\theta}\frac{f(r)}{\sqrt{r}}$ in Eq.
\eqref{kg eq in r coord} and obtain
\begin{equation} \label{kg in btz}
\frac{(r^{2}-r_{+}^{2})}{l^{2}}
\frac{d^{2}f}{dr^{2}}+\frac{2r}{l^{2}}\frac{df}{dr}-
\frac{3f}{4l^{2}}-\frac{r_{+}^{2}f}{4r^{2}l^{2}} 
+\frac{l^{2}\omega^{2}f}{r^{2}-r_{+}^{2}}-\frac{n^{2}f}{r^{2}}-m^{2}f=0 \,.
\end{equation}
To aid further calculations we now go to the tortoise coordinates which are 
effected by the transformation
\begin{eqnarray} \label{tortoise coordinates}
dr_{*} &=& \frac{l^{2}}{r^{2}-r_{+}^{2}}dr, \nonumber \\
r_{*} &=& -\frac{l^{2}}{r_{+}}\coth^{-1}\left(\frac{r}{r_{+}}\right), \nonumber \\
r &=& -r_{+}\coth\left(\frac{r_{+}r_{*}}{l^{2}}\right) \,.
\end{eqnarray}
Note that the horizon $r=r_{+}$ in tortoise coordinates is at $r_*=-\infty$ and spatial infinity
is at $r_*=0$. Computing $\partial_{r}f$ and $\partial_{r}^{2}f$ and substituting in \eqref{kg in btz} we get
\begin{equation} \label{kg in mixed coordinates}
\frac{d^{2}f}{dr_{*}^{2}}-\left(\frac{3r^{2}}{4l^{4}}-\frac{M}{2l^{2}}-
\frac{M^{2}}{4r^{2}}+\frac{n^{2}}{l^{2}}-\frac{Mn^{2}}{r^{2}}+
\frac{m^{2}r^{2}}{l^{2}}-m^{2}M\right)f+\omega^{2}f=0 \,.
\end{equation}
We now transform Eq. \eqref{kg in mixed coordinates} entirely in terms of $r_{*}$
\begin{equation} \label{kg in tortoise btz}
-\frac{d^{2}f}{dr_{*}^{2}}+\frac{1}{\sinh^{2}(\alpha r_{*})}
\left(\frac{3\alpha^{2}}{4}+m^{2}M\right)f
+\frac{1}{\cosh^{2}(\alpha r_{*})}
\left(\frac{\alpha^{2}}{4}+\frac{n^{2}}{l^{2}}\right)f=\omega^{2}f \,,
\end{equation}
where we have used the notation $\alpha=\sqrt{M}/l$
(we also note that the horizon is given by $r_{+}=l\sqrt{M}$, $M$ being the mass of the black hole).
This equation is in the form of standard Schroedinger equation 
\[
-\frac{d^{2}\psi(r_{*})}{dr_{*}^{2}}+V(r_{*})\psi(r_{*})=\omega^{2}\psi(r_{*})
\]
with the potential
\begin{equation} \label{potential in tortoise btz}
V=\frac{1}{\sinh^{2}(\alpha r_{*})}\left(\frac{3\alpha^{2}}{4}+m^{2}M\right)+\frac{1}{\cosh^{2}(\alpha r_{*})}\left(\frac{\alpha^{2}}{4}+\frac{n^{2}}{l^{2}}\right) \,.
\end{equation}
We now need to write the scalar product in terms of the $r_{*}$ coordinate 
and for this we note that in the ansatz $\phi=e^{-i\omega t}e^{in\theta}\frac{f(r)}{\sqrt{r}}$ 
there is an explicit factor of $\sqrt{r}$ sitting in the denominator. The two 
factors of $\phi$ in the scalar product will thus contribute a factor of $r^{-1}$ 
which will cancel the $r$ in the numerator of \eqref{scalar product btz}. Thus, in 
terms of the tortoise coordinates the integration measure is simply $dr_{*}$. 
It is now straight forward to verify that the Laplacian is a self-adjoint operator 
for the boundary condition $\kappa f=df/dr_{*}$ imposed at the horizon $r_{*}=-\infty$. 
This is of the same form as the condition we had in the previous section on flat backgrounds.

Knowing the appropriate boundary condition that makes the Hamiltonian 
operator self-adjoint, we can now go back to the $r$ coordinate in terms of 
which the boundary condition is
\begin{equation} \label{btz boundary condition in terms of r coordinate}
Ag(r)+(r-r_{+})\partial_{r}g(r)=0 \,,
\end{equation}
where $g(r)$ is the solution of the radial part of the Laplacian in the original coordinates. In the above equation both sides are evaluated at $r=r_+$ and $A$ is a constant depending on $\kappa,r_{+},l$. 


Having obtained the conditions for self-adjointness in \eqref{btz boundary condition in terms of r coordinate} the analysis requires
numerical computations. This is not completed because of strong oscillations of solutions near the
horizon for generic boundary values. But we will argue that analysis will lead to similar
results.

Equation \eqref{kg in mixed coordinates} is the Hamiltonian eigenvalue equation in 1-dimension for each value of n.
Comparing this with similar Eq. \eqref{r dependent equation in 2d} we find that the self-adjointness condition
can be satisfied only for finite $n$ up to some maximum value $n_{m}$. The eigenvalues are
negative and $\propto-\kappa^{2}$ since contributions come from the second order derivative term in \eqref{kg in btz}
satisfying boundary conditions. These are close to the boundary. From our flat space computations it is clear that the number of
bound states is finite and proportional to $r_{+}$ $(r_{+}^{2})$ in two dimensions (three dimensions). The same result
continues to be true even in the BTZ case as can be seen by
near horizon analysis of solutions of \eqref{kg in btz} satisfying the boundary conditions \eqref{btz boundary condition in terms of r coordinate},
though obtaining the exact number of solutions looks difficult.
The reason for this is that in the Hamiltonian \eqref{kg in mixed coordinates}, we see that for large $r$ 
there is a term in the potential which is proportional to $r^2$ like the harmonic
oscillator. Hence, the spectrum has eigenvalues which are positive and wave functions are
supported over a larger length scale controlled by cosmological constant. We should separate the two
kinds of states, and this can be analyzed in the large $r_+\gg l\gg 1/\kappa$
regime. This problem is akin to self-adjointness for harmonic oscillator on $\mathbb{R}^{2}-\mathbb{D}$. 
Furthermore the boundary (horizon), which is at $r_{*}=-\infty$, is actually located at a finite
distance $r_{+}$ in usual coordinates. Hence the self-adjointness condition is related to
the dilatation operator $(r-r_+)\partial_{r}$ near the horizon. The scaling operator along with the Hamiltonian forms
elements of an SL(2,R) algebra which has been used through conformal quantum mechanics to understand the near horizon
dynamics \cite{camblong, kumargupta1}.

\section{Discussion}
In this paper we find novel bound state solutions for Laplacian on manifolds 
with boundary. These arise by requiring it to be  self-adjoint. These states 
are localized close to the boundary and serve to explain the states of a 
black hole contributing to the entropy. They also contribute to Hawking radiation
which will be explored later. These features arise through brickwall mechanism
and quantum mechanical origin justifying the boundary conditions on the horizon.
The horizon states can be compared with membrane paradigm also \cite{damour, thorne}. 
They can also be looked as states obtained from spin network of LQG though our states 
contribute directly to energy whereas spin network give area bits.

In this connection we wish to point out an interesting result from \cite{ibort}.
Here a general analysis is done for self-adjoint extensions and the most general 
boundary conditions are characterized by an infinite dimensional 
 unitary matrix linking the boundary data.
It is given by 
\begin{equation}
(\phi+i\partial_{\overrightarrow{n}}\phi)=U(\phi-i\partial_{\overrightarrow{n}}\phi)
\end{equation}
evaluated at the boundary and $\overrightarrow{n}$ stands for the normal. They also point out that 
if $U$ has an eigenvalue $-1$ then the extensions given by $Ue^{-it}$
have a negative energy state which is an edge state. They are characterized 
by the length parameter. However, when the Hamiltonian has in addition 
potential term which contributes 
positive energy to the state which increases with $n$,  
then there is a balance of these
contributions resulting in finite number of states. 
This results in the number of edge 
states being constrained by the radius.

In the case of Schwarzschild black hole the situation is different 
since it is asymptotically flat. This work is in progress and will be presented 
elsewhere. 

\section*{Acknowledgements}
Authors thank A.P. Balachandran, K.S. Gupta, Parameswaran Nair and Michael Berry for various discussions
on self-adjoint operators and boundary conditions. TRG thanks Robert Wald for 
discussions on black holes and boundary conditions on the horizon.

TRG acknowldges the support of A. Ferraz, Director, IIP, Natal, Brazil
where this work was completed.

\end{document}